\documentclass[pre,aps,showpacs,twocolumn,preprintnumbers,amsmath,amssymb,superscriptaddress,showkeys]{revtex4-1}

\usepackage{epsfig}

\usepackage{graphics}
\usepackage{dcolumn}
\usepackage{bm}

\usepackage{color}

\begin{document}


\title{Mesoscopic model for DNA G-quadruplex unfolding}

\author{A. E. Bergues-Pupo}

\affiliation{Dpto. de F\'isica de la Materia Condensada,
Universidad de Zaragoza. 50009 Zaragoza, Spain}

\affiliation{Instituto de Biocomputaci\'on y F\'isica de Sistemas Complejos (BIFI), Universidad de Zaragoza, 50009 Zaragoza, Spain}

\author{I. Guti\'errez}

\affiliation{Instituto Madrile\~{n}o de Estudios Avanzados en Nanociencia,
C/Faraday 9, Cantoblanco, 28049 Madrid, Spain}

\author{J.~R. Arias-Gonzalez}

\affiliation{Instituto Madrile\~{n}o de Estudios Avanzados en Nanociencia,
C/Faraday 9, Cantoblanco, 28049 Madrid, Spain}

\affiliation{CNB-CSIC-IMDEA Nanociencia Associated Unit
``Unidad de Nanobiotecnolog\'{i}a", Cantoblanco, 28049 Madrid, Spain}

\author{F. Falo}

\affiliation{Dpto. de F\'isica de la Materia Condensada,
Universidad de Zaragoza. 50009 Zaragoza, Spain}

\affiliation{Instituto de Biocomputaci\'on y F\'isica de Sistemas Complejos (BIFI), Universidad de Zaragoza, 50009 Zaragoza, Spain}

\author{A. Fiasconaro}

\affiliation{Dpto. de F\'isica de la Materia Condensada,
Universidad de Zaragoza. 50009 Zaragoza, Spain}

\date{\today}

\begin{abstract}
Genomes contain rare guanine-rich sequences capable of assembling into four-stranded helical structures, termed G-quadruplexes, with potential roles in gene regulation and chromosome stability. Their mechanical unfolding has only been reported to date by all-atom simulations, which cannot dissect the major physical interactions responsible for their cohesion. Here, we propose a mesoscopic model to describe both the mechanical and thermal stability of DNA G-quadruplexes, where each nucleotide of the structure, as well as each central cation located at the inner channel, is mapped onto a single bead. In this framework we are able to simulate loading rates similar to the experimental ones, which are not reachable in simulations with atomistic resolution. In this regard, we present single-molecule force-induced unfolding experiments by a high-resolution optical tweezers on a DNA telomeric sequence capable of forming a G-quadruplex conformation. Fitting the parameters of the model to the experiments we find a correct prediction of the rupture-force kinetics and a good agreement with previous near equilibrium measurements. Since G-quadruplex unfolding dynamics is halfway in complexity between secondary nucleic acids and tertiary protein structures, our model entails a nanoscale paradigm for non-equilibrium processes in the cell.
\end{abstract}


\maketitle

\section{Introduction}

G-Quadruplexes (G4) are non-canonical conformations of DNA or RNA sequences rich in Guanine nucleobases. Unlike the classic double helix~\cite{Arias2014}, the basic structural unit in the G4 motif is the \textit{G-tetrad}, a planar arrangement of four guanines (G) nucleobases held by Hoogsten hydrogen bonds \cite{Burge2006}. The presence of monovalent cations like K$^{\rm{+}}$ and Na$^{\rm{+}}$ stabilizes the highly electronegative central channel along the axis of the G4 stem. G4s have been observed in some biological sequences within telomeres and in promoter regions \cite{Lam2013,Siddiqui2002}, being their high mechanical stability one relevant property in their potential role in processes such as gene expression and chromosome maintenance \cite{endoh2016mechanical,Hansel2017}.

The mechanical stability of G4s have been studied at the single molecule level by means of AFM and optical tweezers (OT)~\cite{Mes2012,Mao2012,Long2013,Ghimire2014,Garavis2013}. Mechanical stability in this framework is defined in terms of the unfolding force or unfolding free energy. Typical values of the unfolding forces for G4 range between 20-30 pN and depend on the G4 topology and the type of ion present in the G4 channel \cite{Mes2012}.  Single-molecule characterization is at present limited by the temporal and spatial resolutions of the experimental techniques. Complementary information as the structural changes during the unfolding is obtained by molecular simulations.


The dynamics of the G4 has been modeled at different temporal and length scales, from quantum calculations \cite{Yurenko2014,Poudel2016} and molecular dynamics simulations \cite{Li2010,Islam2013,Stadlbauer2013,Li2009,Yang2011,Pupo2015} to mesoscopic approaches \cite{linak2011moving,rebic2014multiscale}. Molecular dynamics allows an atomistic description of G4 properties.
In this regard, we previously studied the mechanical unfolding of a fragment of the human telomeric sequence that can be folded in different geometries by using Steered Molecular Dynamics. We showed that the unfolding pattern in the force-extension curves is correlated with the loss of coordination of the central ions in the G4 and that its stability is significantly decreased if the ions are removed~\cite{Pupo2015}. These results cannot be compared directly with the experimental results due to the high pulling velocity used in this molecular dynamics simulation (around 6 orders of magnitude higher than in the experiments), which is known to affect the unfolding forces.

\begin{figure*}[!t]
\begin{center}
\includegraphics[width=0.23\textwidth]{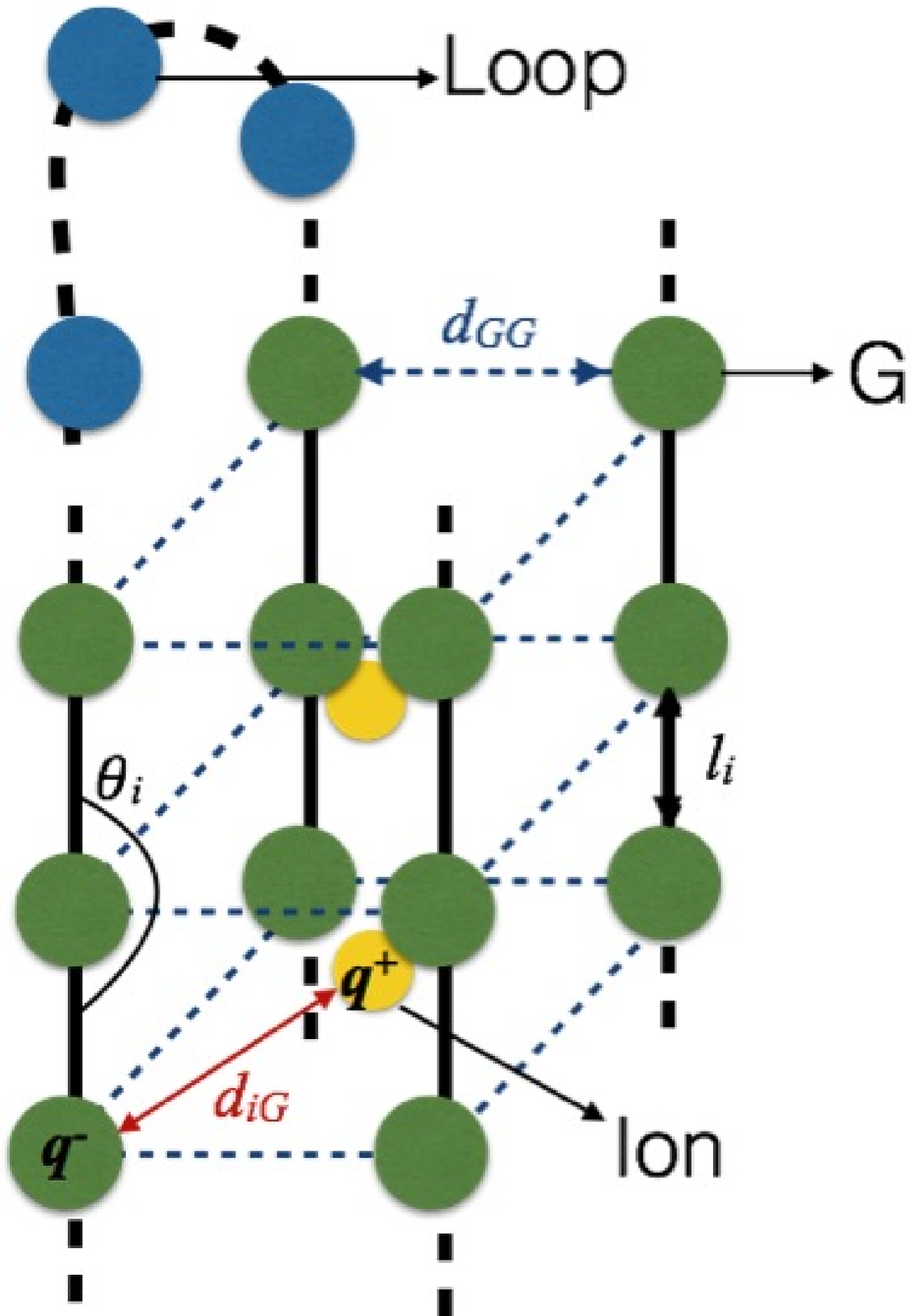}
\includegraphics[width=0.55\textwidth]{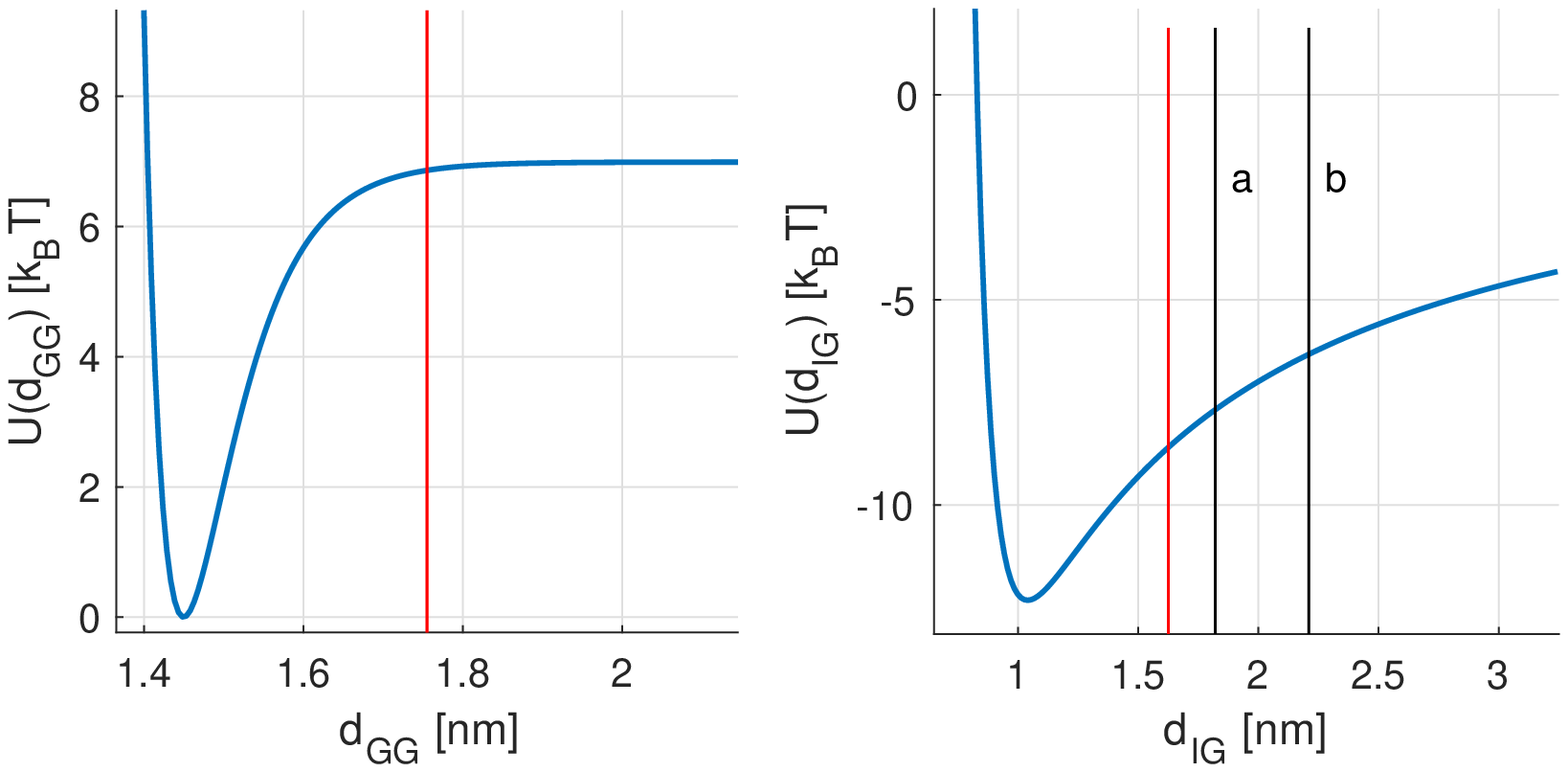}
\end{center}
\caption{Mesoscopic model for G4. \textit{Left:} Scheme of a parallel G4 assembly, taken as exemplary structure, where each nucleotide is represented by a single bead (not to scale). For simplicity, the twist between successive G-tetrad planes is not represented. Such rotation makes the distance between two consecutive planes ($p_0=0.5\, l_0$, see the text for details) lower than the equilibrium distance between two consecutive beads in the chain ($l_0=0.65\,\rm{nm}$). \textit{Right:} Potential between Guanines of the same plane $U(d_{GG})$ and between one ion and its neighbor Guanines $U(d_{IG})$. The red lines mark the  threshold distances used to define the coordination numbers between Guanines and between the central ions and the neighbor Guanines. The black lines in the panel of $U(d_{IG})$ delimit the interval [a,b] where the coulomb force attraction is linearly shifted to zero.}
\label{fig:model_G4}
\end{figure*}

Larger temporal and spatial scales can be achieved by means of mesoscopic models. Unlike for double-stranded (ds) DNA or single stranded (ss) structures, like DNA hairpins, only few mesoscopic models have been developed for G4 \cite{linak2011moving,rebic2014multiscale,2016Stadlbauer} and none, to our knowledge, to characterize the mechanical unfolding. Margaret \textit{et al.} \cite{linak2011moving} developed a bead-spring model for dsDNA with three beads per nucleotide. They studied thermal properties of dsDNA and also those of ssDNA, specifically the melting of a DNA hairpin and the folding of the thrombin aptamer (a G4 with two G-tetrad planes). Rebic \textit{et al.} developed a mesoscopic model following a \textit{bottom up} approach \cite{rebic2014multiscale}. Their model presents three different beads: one bead for the Guanines, another for the nucleotides in the loop, and the last one for the ions, which interact by means of tabulated potentials.

In this work we propose a mesoscopic model for a fragment of the G4 human telomeric DNA with a resolution of one bead per base and investigate both its thermal and its mechanical stability. The simplified representation of the model allows, on the one hand, the exploration of the mechanical unfolding at lower velocities than permitted in atomistic simulations and, on the other hand, the direct comparison with a high resolution mechanical unfolding experiment we herein present.

\section{Methods}
\subsection{DNA G4 model}
Each nucleotide and the two central ions are represented by a single bead as depicted in fig.~\ref{fig:model_G4} where $l_i$ is the distance between two consecutive beads in the chain; $d_{GG}$ is the distance between two Guanines of the same plane; $d_{IG}$ is the distance between each central ion and its neighbor Guanines (this distance is defined for the eight closest Guanines to each ion) and $\theta_i$ is the angle between three consecutive Guanines. The Hamiltonian of such system is composed by the following terms.

\begin{itemize}
\item A harmonic interaction between two consecutive beads $U_{str}=\sum_i \frac{1}{2}k_s(l_i-l_0)^2$, where $l_i=|\textbf{r}_{i+1}-\textbf{r}_{i}|$, $\textbf{r}_{i}$ is the
position vector of the $i$-th particle, $l_0$ is the equilibrium separation between beads and $k_s$ the elastic constant of the chain. We take $l_0=0.65 \,\rm{nm}$, which is approximately the distance between consecutive phosphor backbone atoms in the crystal structure of the human telomeric parallel G4 \cite{2002Parkinson}.

\item A Morse potential between consecutive beads of the same plane to simulate the Hoogsten hydrogen bonds between the Guanines: $U_{GG}=\sum_p \sum_g D_G ({\rm e}^{-\alpha_G (d_{GGg}-r_0)}-1)^2$, where the sums go over the number of planes $p=1,...,3$ and the number of Guanines in each plane $g=1,...,4$, $d_{GGg}$ is the distance between two consecutive Guanines of the same plane and $r_0$ the equilibrium length of the side of each plane. The strength of the hydrogen bond interaction highly depends on their environment. In the case of the G4, it has been shown that the hydrogen bonds between Guanines are stronger when all the bonds of the same plane are formed \cite{Yurenko2014}. To take into account this cooperative effect, we take the depth of the Morse potential $D_G$ to be dependent on the distances between the Guanines of the same plane as follows:
\begin{equation}
D_G=D_0 \lbrace 1+2e^{-\delta (\sum_g d_{GGg}-4r_0)} \rbrace,
\label{eq:eqn_morse}
\end{equation}
where $d_{GGg}$ are the four distances in the same plane and $\delta$ sets the length scale for the decay of $D_G$. Thus $D_G$ varies from $D_G=3D_0$ when the G4 is folded and the side of the plane is $r_0$, to $D_0$ when any of the distances $d_{GGg}$ increases beyond $\delta^{-1}$.

\item An interaction potential between each central ion and its neighbor Guanines $U_{IG}=\sum_i \sum_g A/d_{ig}^{12}-Q_{ig}/d_{ig}$, where the sums go over the two ions $i=1,2$ and the eight neighbor Guanines $g=1,...,8$,  $A/d_{ig}^{12}$ is a repulsive term that accounts for the excluded volume effect and $Q_{ig}$ describes the strength of the \textit{effective} attraction between the ion and the Guanine. The constant $A$ is selected in such a way that the minimum of $U_{IG}$ is located at $\sqrt{2(r_0/2)^2+(p_0/2)^2}$ ($p_0$ is the distance between the centers of two consecutive planes). The Coulomb contribution $Q_{ig}/d_{ig}$ is shifted linearly to zero in the interval $a<b$ indicated in figure~\ref{fig:model_G4} with the two black vertical lines. A purely repulsive interaction between the two ions $U_{I_1I_2}=Q_{I_1I_2}/d_{I_1I_2}$ is also included. We set  $Q_{ig}=2.5Q_{I_1I_2}$. It is worth noting that even if the interactions between the two ions and between each ion and the Guanine have an electrostatic origin, their features are different. In fact, the interaction between the Guanine base is due to the metal-ion coordination between the ion and the oxygen O6 of the Guanine, and then is more sensitive to their distance separation than if it were a pure electrostatic interaction~\cite{2016Bhattacharya}. For this reason we introduce, as a typical procedure adopted in these cases, a cutoff distance for the interaction between the ion and the Guanines.

\item A bending energy interaction between the three consecutive beads that belong to each side of the G4 stem  $U_{ben}=\sum_i k_b/2\left[ 1-\cos(\theta_i-\theta_0) \right]$, where $k_b$ is a bending elastic constant, $\theta_0 \approx \, 150^o$ due to the twist between planes (not represented in Fig.~\ref{fig:model_G4}) and $\theta_i$ is the angle
between vectors $\textbf{l}_{i+1}$ and $\textbf{l}_{i}$. This term accounts for the stacking interactions between the consecutive Guanines and confers stability to the G4 stem. For the beads of the loops the bending can be neglected.

\item A repulsive Lennard Jones interaction $V_{\rm LJ}=\sum_i \sum_{j>i}4\epsilon \left[ (\sigma/r_{ij})^{12}-(\sigma/r_{ij})^6 \right]$ if $r_{ij}<1.122 \sigma$ between all the beads of the G4. This term accounts for the excluded volume effect.  \end{itemize}.

The dynamics of the system is obtained from the overdamped equations of motion of the $j$-th bead of the G4 ($j=1,...,21$)
\begin{eqnarray}
m_j\dot{\textbf{r}}_{j}=-\nabla_{\textbf{r}_j}U_{str}-\nabla_{\textbf{r}_j}U_{GG}-\nabla_{\textbf{r}_j}U_{ig}-\nabla_{\textbf{r}_j}U_{ben} \nonumber \\-\nabla_{\textbf{r}_j}V_{LJ}+\sqrt{2k_BTm_j} \mathbf{\eta}_j(t),
\label{eq:mov1}
\end{eqnarray}
and for the $i$-th bead representing the ions ($i=1,2$)
\begin{equation}
m_i\dot{\textbf{I}}_{i}=-\nabla_{\textbf{I}_i}U_{ig}-\nabla_{\textbf{I}_i}U_{I_1I_2}+\sqrt{2k_BTm_i} \eta_i(t).
\label{eq:mov2}
\end{equation}

The last terms in eqs. \ref{eq:mov1},\ref{eq:mov2} represent the thermal contribution as a Gaussian uncorrelated noise. The damping is taken implicit into the time units. The mass of the ion $m_i$ and of the nucleotide $m_j$ are taken as $m_j=3.85\, m_i$. We use the following dimensionless units: length is given in units of $l_0=0.65\,\rm {nm}$ and energy in units of $D_0$. The energy and time units are derived in the next sections in order to match the experimental values of the G4 melting temperature and unfolding force with the simulations. The rest of the parameters in the dimensionless units are $D_0=1$, $k_s=100$, $k_b=2$, $\alpha_G=10$ and $\delta=0.3$.

Systems of equations \ref{eq:mov1} and \ref{eq:mov2} are integrated with the stochastic Euler algorithm with $dt=10^{-4}$. For the melting simulations, were the dynamics is studied at different temperatures, the simulations are started from the lowest temperature and a folded conformation of the G4. The final positions and velocities at each temperature are used as the initial conditions for the next temperature. Each simulation lasts for $10^7$ time steps, from which, the first $10^6$ steps are for thermalization. Pulling simulations are conducted at a temperature lower than the melting until the G4 is unfolded.

\subsection{Force-induced unfolding experiments}
\begin{figure}[!th]
\begin{center}
\includegraphics[width=0.37\textwidth]{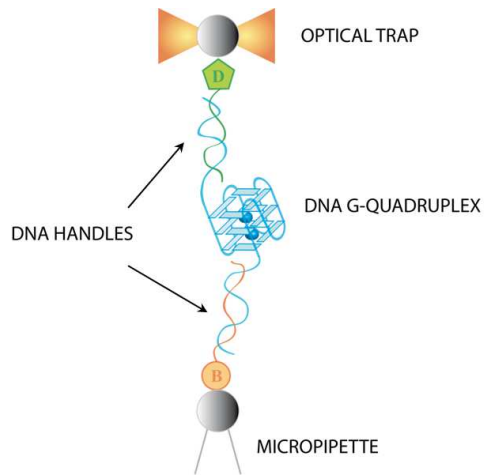}
\end{center}
\caption{Experimental configuration for the mechanical unfolding of the G4 at the single-molecule level. The two ends of the G4 are tethered by two dsDNA handles, which are in turn attached to micronsized beads via biotin-streptavidin and digoxigenin-antidigoxigenin bonds. One of the beads is kept fixed by air suction to a micropipete while the other one is trapped in the laser-beam focus forming the OT. ``B" stands for biotin and ``D" for digoxigenin.}
\label{fig:Experiment}
\end{figure}

To adjust and validate the mesoscopic model, we performed constant velocity pulling experiments in which a DNA telomeric sequence that yields a G4 is unfolded by means of a high-resolution OT device, as depicted in Fig.~\ref{fig:Experiment}. The central hexanucleotide-repeat sequence is flanked by dsDNA handles for its manipulation in the optical setup (see below). The trap stiffness is $k= 0.135 \pm 0.004 \,\rm{pN/nm}$. The micropipette is moved relative to the optically trapped bead with a pulling velocity $v= 11.8 \pm 1.4 \,\rm{nm/s}$ near the rupture event. The elastic constant acting on the G4 due to the dsDNA handles is estimated from the slope of their force-extension curve in the enthalpic elasticity regime before the rupture, which gives $k_{DNA} \approx 0.4 \pm 0.1 \,\rm{pN/nm}$.

\subsubsection{Synthesis of telomeric DNA molecules}
The molecular construction for OT experiments consists of a telomeric G4-forming sequence (5'-TATA (GGGTTA)$_5$ TAGT-3') flanked by two dsDNA handles, one of 650 bp at the 5' end (handle \textit{A}) and the other of 401 bp at the 3' end (handle \textit{B}), for their attachment to beads through digoxigenin-antidigoxigenin and biotin-streptavidin labeling, respectively (Fig.~\ref{fig:Experiment}). The extra repetition and the flanking ssDNA nucleotides provide configurational flexibility for the formation of the G4 in the presence of the dsDNA handles. The three DNA fragments were obtained by PCR amplifications of a conveniently modified pUC18 plasmid~\cite{Garavis2013}. Handle \textit{B} was labeled during its PCR amplification using a 5'-biotinylated primer (Integrated DNA Technologies, IDT, Coralville, IA) and handle \textit{A} was modified afterwards adding a very short tail of 2-3 digoxigenin-dUTP at its 3' end (DIG Oligonucleotide Tailing Kit, 2nd Generation (Roche); 37$^o$C-15 min). Both DNA templates and labeled handles were purified using Wizard\textregistered ~SV Gel and PCR Clean-Up System (Promega). Finally, equimolar amounts of the telomeric DNA and the two DNA labeled handles were mixed and annealed in the presence of KCl in a PCR apparatus using the annealing procedure described in~\cite{Garavis2013}.

\subsubsection{Optical setup}
Measurements have been performed in a dual-beam optical trap in which two diode lasers (250 mW at maximum power, 808 nm wavelength) and associated optics are compacted into a miniaturized instrument suspended from the ceiling~\cite{2015deLorenzo}. Each beam is delivered by an optical fiber and its position in the plane perpendicular to the optical axis is controlled by bending the optical fibers using piezoelectric crystals. The two laser beams are counter-propagating and brought to the same focus with orthogonal polarizations, which allow their optical paths to be separated using polarizing beam splitters. Each beam is passed through a pellicle beam-splitter that redirects about 5\% of the intensity, which is used to measure the position of the trap. The remaining light is focused through water-immersion objectives (NA=1.20) to form the optical trap in a microfluidics chamber, which also contains a micropipette. The light exiting the trap from each beam is collected by the opposite objective lens, which redirects it to position-sensitive detectors that monitor the three force components acting on the trapped bead. Force is measured using the light momentum conservation~\cite{2003Smith}. This setup design reduces the mechanical drift and allows a large measurement stability over time thus enhancing the discernibility of the rupture events associated with the unfolding of the DNA structure. It allows an approximate force resolution of 0.1 pN and a distance resolution of 0.5 nm.

\section{Results}
\begin{figure*}[!htb]
\begin{center}
\includegraphics[width=0.5\textwidth,angle=-90]{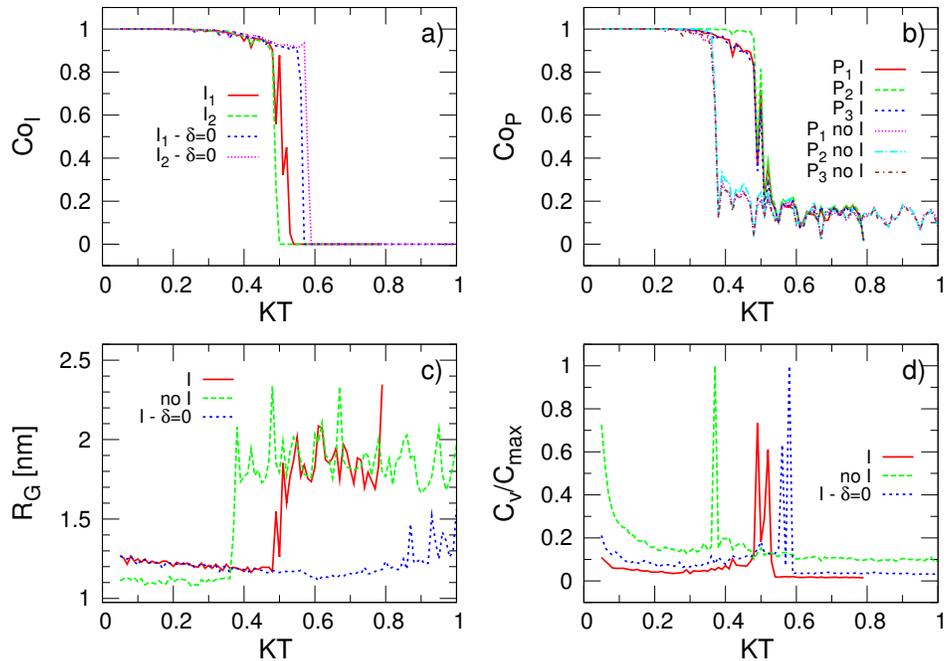}
\end{center}
\vspace{5mm}
\caption{Melting curves in terms of different magnitudes as a function of the thermal energy $KT$: a) Average coordination number of each ion $Co_I$. $I_1$ and $I_2$ stand for each of the ions, whereas $\delta=0$ indicates the simulations without coperativity in the Morse potential term of Eq.~\ref{eq:eqn_morse}). b) Average coordination number in each plane $Co_P$ with (I) and without (no I) the central ions. c) Gyration radius $R_G$ with and without the central ions. d) Heat capacity at constant volume $C_v$ with and without the central ions. Each of the above magnitudes are defined in the text. $KT$ represents the thermal energy in dimensionless units. The energy units are determined next according to the experimental melting temperature.}
\label{fig:Cp1}
\end{figure*} 

\subsection{Melting simulations}
In this section we study the thermal stability of the G4 with our model. To this end, different magnitudes are calculated as a function of the temperature: the average coordination number of each ion $Co_I$, the average coordination number of each plane $Co_P$, the radius of gyration of the molecule $R_G$ and the heat capacity $C_v$. The coordination number of each ion is 1 if the distance between it and its neighbor Guanine is lower than $1.63 \,\rm{nm}$ ($2.5 l_0$) and 0 otherwise. In each plane the coordination number between two consecutive Guanines is put equal to 1 if the distance between them is lower than $1.75 \,\rm{nm}$ (a distance around the plateau of the Morse potential is reached) and 0 otherwise. $Co_I$ and $Co_P$ are defined as the average over the total possible contacts for each ion-Guanine (8) and in each plane (4), respectively and over the simulation time at each temperature.

\begin{figure}[!ht]
\begin{center}
\includegraphics[width=0.47\textwidth]{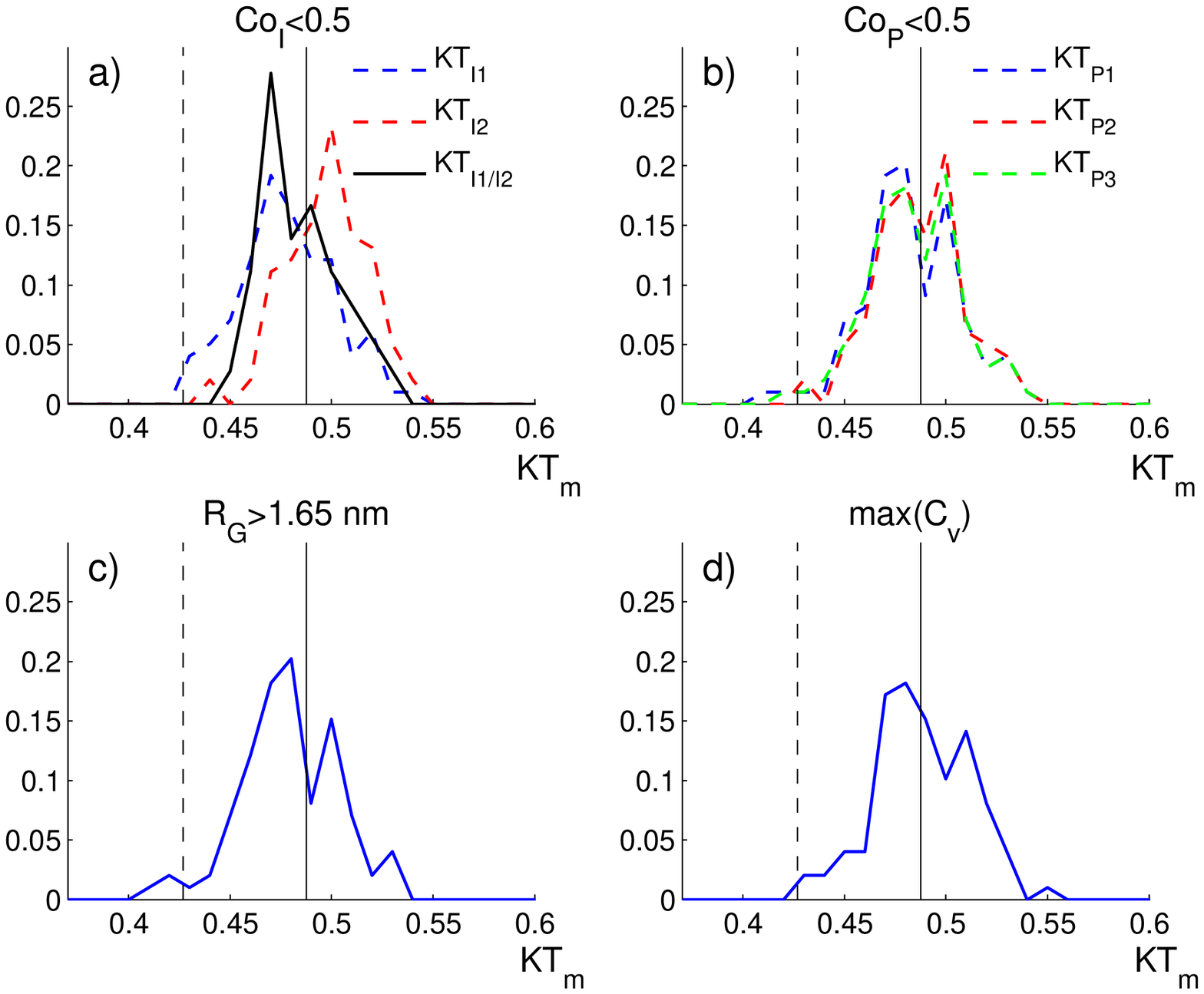}
\end{center}
\caption{Distribution of thermal energy at the melting temperature  calculated from different magnitudes over 100 realizations. a) Temperature at which the coordination of the ions $Co_I$ goes below 0.5; b) Temperature at which the coordination of the planes $Co_P$ goes below 0.5; (c) Temperature at which the radius of gyration $R_G$ goes over $1.2 \,\rm{nm}$; and d) Temperature at which $C_v$ takes the maximum value. The melting temperature value for defining our energy unit is $KT_m^{\dagger}=0.4875$ (solid line) and the room temperature, at which the pulling simulations are performed is $KT=0.4269$ (dashed line).}
\label{fig:T_dist}
\end{figure}

Figure~\ref{fig:Cp1} shows the melting curves of the G4 described by the above mentioned magnitudes. For comparison, the results of a simulation without the central ions are also included. In both panels a) $Co_I$ and b) $Co_P$ we observe that the curves decrease in a stepwise fashion approximately at the same temperature for the two measures. Analogously, in the presence of the ions, $R_G$ (panel c)), and $C_V$ (panel d)) show a similar stepwise change at the same temperature. The narrow temperature interval in which $Co_P$ goes to zero, as well as the sharp transition of $R_G$, are a consequence of the coperativity term included in the definition of the Morse potential (Eq.~\ref{eq:eqn_morse}).
Conversely, in absence of the coperativity term ($\delta=0$), the measures undergo their changes at higher temperatures (the breaking of the planes $Co_P$ is not shown for simplicity): in the case of the gyration radius the transition is quite smooth and the dissociation of the three planes $Co_P$ takes also place in a wide interval instead of appearing as sharp transitions. The transition temperatures are then very different from each other and do not permit any definition of a melting temperature. In other terms, the contribution of the cooperative term appears essential in determining the steepness and so the uniqueness of the melting temperature in the thermal unfolding.

Moreover, the abrupt changes in $Co_P$, $R_G$ and $C_v$, characteristic of the melting, is highly influenced by the presence of central ions. We observe in Fig.~\ref{fig:Cp1} (panels b), c), and d)) that the transitions of all the measures occur at a lower temperature if the ions are not present (`no I' labels), indicating that their coordination increases the thermal stability of the G4~\cite{Pupo2015}.

It is important to note that the sharp changes in the behavior of all the curves of the magnitudes described above occur at the same temperature, demonstrating the robustness of the model. It gives the possibility to use indistinctly any of those magnitudes to quantify the thermal unfolding and define the melting temperature $KT_m^{\dagger}$. In each trajectory, we use the mean of the two temperatures $KT_{I1}$ and $KT_{I2}$ at which the coordination $Co_I$ of ion 1 and 2 respectively drop below the value 0.5, namely $KT_{I1/I2}$. To account for the stochastic effects in the unfolding, we repeated the melting simulations $N_s=100$ times from which we set $KT_m^{\dagger}=\frac{1}{2}(\langle KT_{I1} \rangle+\left<KT_{I2}\right>) =\frac{1}{2}\langle KT_{I1/I2} \rangle$, where $\langle\cdot \rangle$ denotes the ensemble average. The distribution of melting temperatures is presented in figure~\ref{fig:T_dist} (panel a). The other panels of the figure present analogue distributions by using magnitudes different than $C_{O_I}$: the loss of the plane coordination $Co_p$ (panel b), the increase of the gyration radius (panel c), and the position of the peak in the heat capacity $\max\{C_v\}$ (panel d). All the measures appear equivalent. By using the ion coordination $Co_I$, the resulting melting temperature is $KT_m^{\dagger}=0.4875$, represented by a solid line in each panel of figure~\ref{fig:T_dist}.

We can now adjust the energy unit of the model. Taking the experimental value $T^*_m=65 \,\rm{^{\circ}C}$ of the melting temperature of a telomeric sequence in [K$^{\rm{+}}$] solution reported in \cite{Phan1998} we get: $E_u=D_0k_BT^*_m/KT_m^{\dagger}=2.33\,k_BT$, ($T=298 \,\rm K$). The unit of force is $F_u=E_u/l_u=14.7 \,\rm{pN}$. In the next section, the pulling simulations are conducted at $KT=0.4269$ which corresponds to $T=22.8 \,\rm{^{\circ}C}$ in real units. This temperature value is represented with dashed line in figure \ref{fig:T_dist}. Note that this temperature value is inside the distribution of melting temperatures. Thus, when doing the pulling at this temperature, there is a non-negligible probability that the unfolding occurs due to thermal effects.

\subsection{Mechanical unfolding at physiological conditions}

\begin{figure}[!b]
\begin{center}
\includegraphics[width=0.3\textwidth]{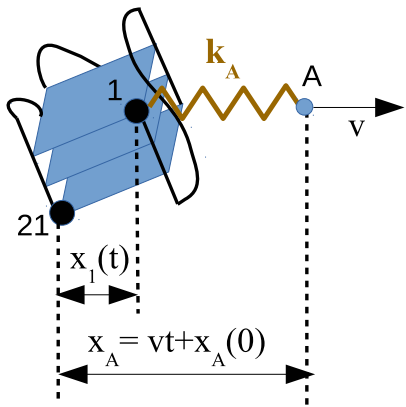}
\end{center}
\caption{Scheme of the pulling simulations. The first and last nucleotides in the G4 chain are represented by  black beads and labeled as 1 and 21, respectively. Bead 21 is fixed to the origin of coordinates, whereas bead 1 is attached to a harmonic spring $k_A$ whose end A is displaced at a constant velocity $v$. The distance projections along the pulling direction $x_A$ and $x_1$ are calculated during the simulations.}
\label{fig:F_pulling}
\end{figure}

In a previous work we studied the mechanical unfolding of different G4 conformations at the atomistic level by means of steered molecular dynamics and showed that the force measured during the unfolding was correlated with the loss of coordination of the central ions \cite{Pupo2015}. In that work, a harmonic spring $k_A$ is attached to one atom of the extreme of the G4 and displaced at constant velocity $v$, while another atom at the opposite end is fixed.
The component of the force along the pulling direction is calculated as $F=k_A(x_A(t)-x_1(t))$, where $x_A(t)$ and $x_1(t)$ are the components of the distance along the pulling direction of the spring end (point A) and the pulled bead 1, respectively, as depicted in Fig. \ref{fig:F_pulling}. The bead 21 is fixed to the origin of coordinates.

\begin{figure}[!th]
\begin{center}
\includegraphics[width=0.31\textwidth,angle=-90]{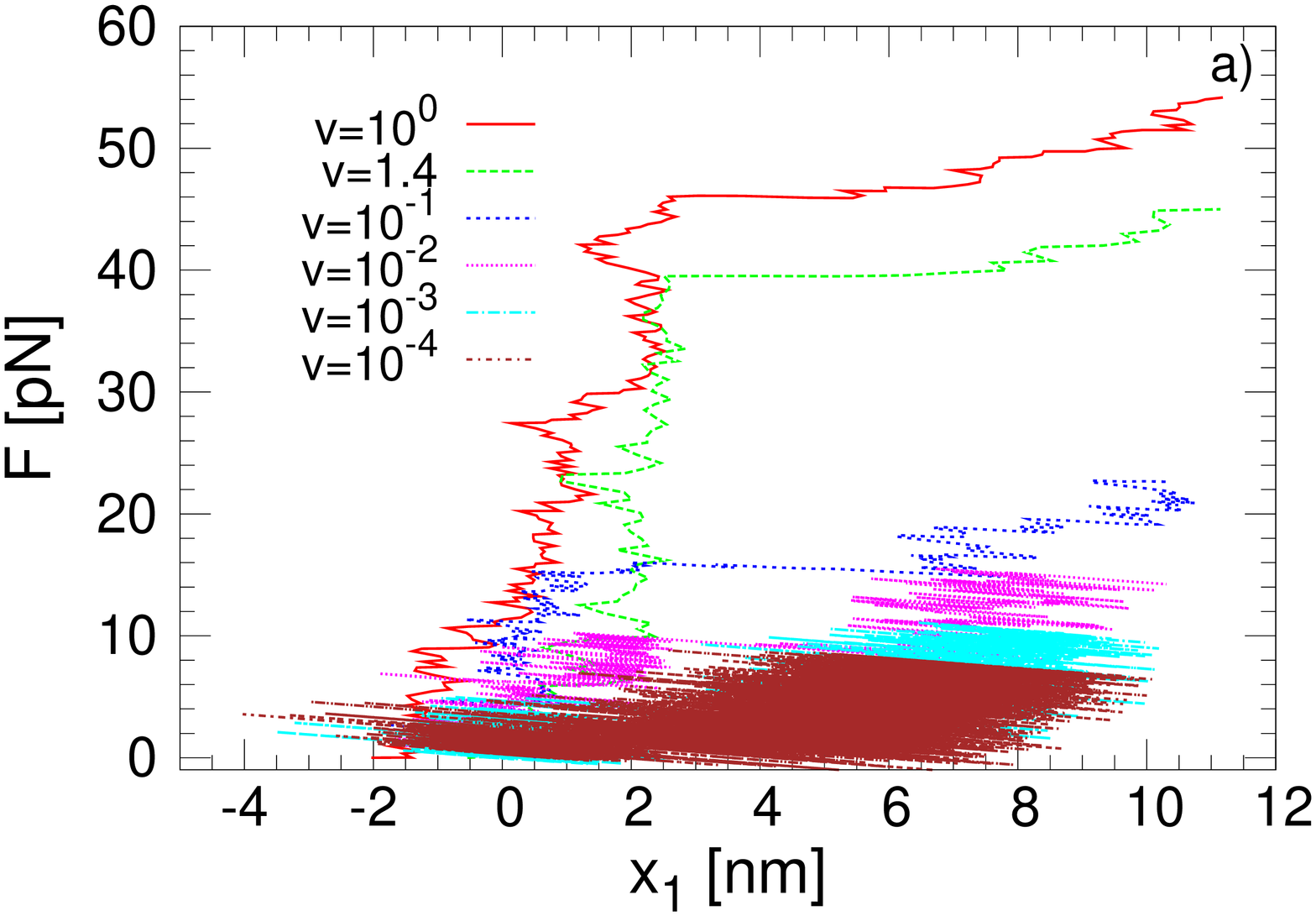}
\includegraphics[width=0.31\textwidth,angle=-90]{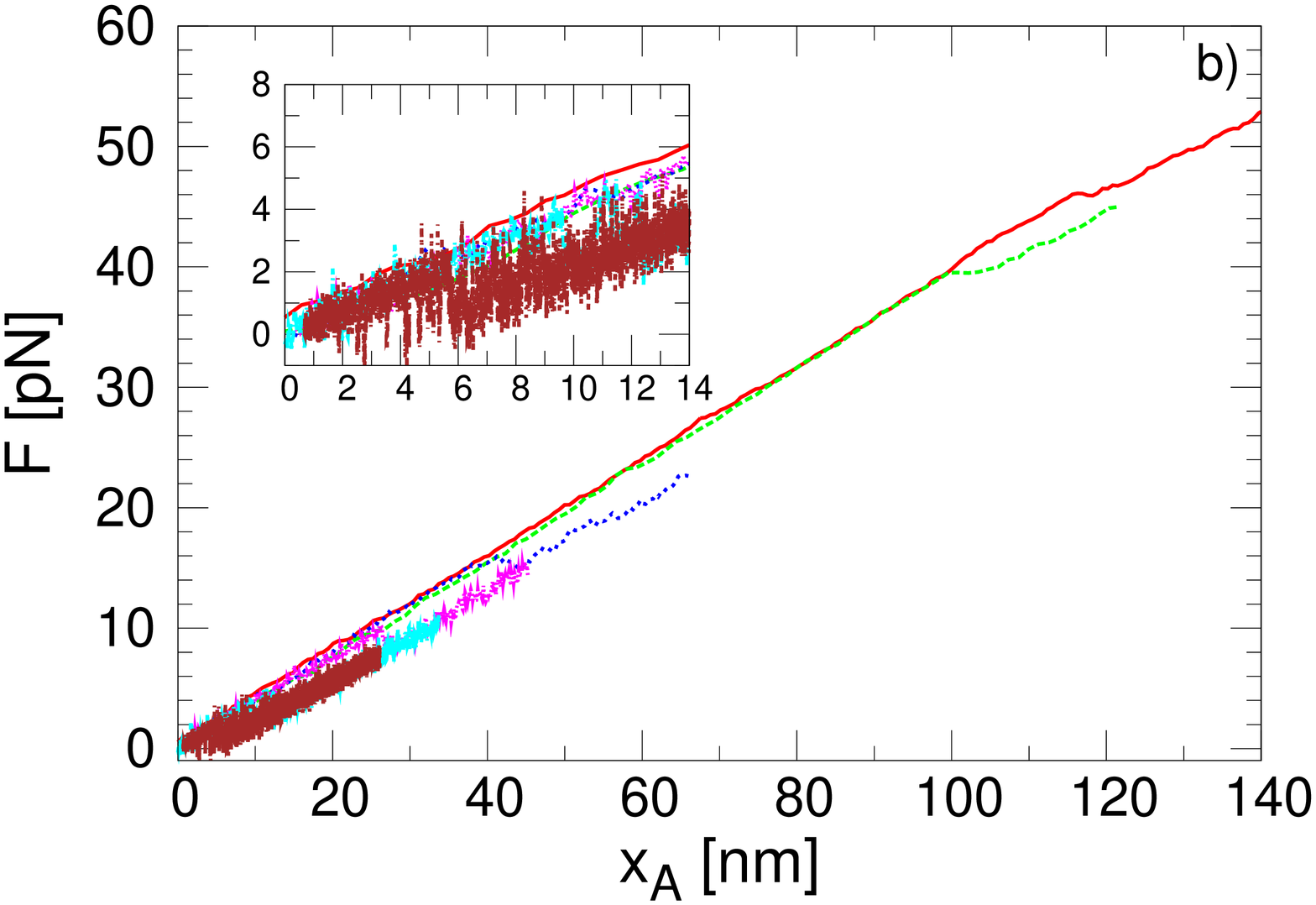}
\includegraphics[width=0.31\textwidth,angle=-90]{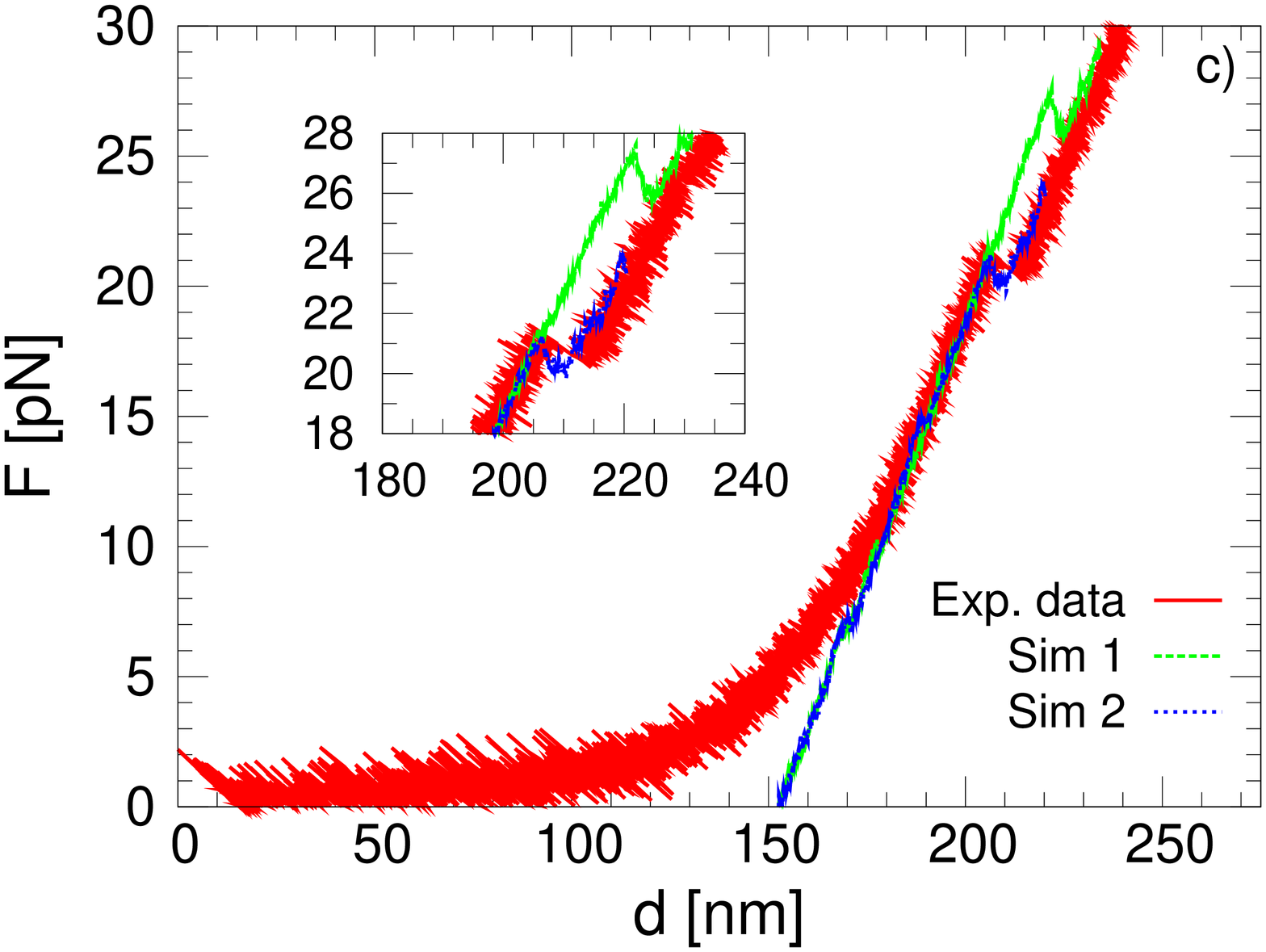}
\end{center}
\caption{a) Force as a function of the extension $x_1$ with the mesoscopic model at different pulling velocities. b) Force as a function of the position of the pulling spring $x_A$ with the mesoscopic model at different pulling velocities. The velocity is given in dimensionless units. c) Comparison of the force-extension curve between two simulations at $v=0.08$ and one experimental realization of the mechanical unfolding. The experimental curve shows superimposed the entropic elasticity regime of the double-stranded DNA handles at low forces (see the Experimental Section).}
\label{fig:F_single_ch}
\end{figure}

The unfolding force value $F_u$ obtained in the all-atom simulations is in the order of $10^2 \,\rm{pN}$  \cite{Pupo2015}, one order of magnitude higher than previous experiments of G4 mechanical unfolding \cite{Mao2012,Garavis2013} and the experiment we present here where $F_u \approx 20.4 \pm 6.9$~pN (mean $\pm$ s.d., N=163 experiments). This is due to the high values of both the parameters velocity $v=1 \,\rm{nm/ns}$ and elastic constant $k_A=1650 \,\rm{pN/nm}$ necessary for all-atom simulations in order to reach a reasonable simulation times. With our mesoscopic model, we are able to decrease both values and to obtain unfolding forces comparable with the experimental ones. The elastic constant used in the mesoscopic simulations is set to $k_A=0.4\,\rm{pN/nm}$, in accordance to our experimental value, as explained in the previous section.

Figures~\ref{fig:F_single_ch}a), and b) show the force $F$ as a function of the extension $x_1$ and the position of the spring $x_A$ for the different pulling velocities, respectively.  All the curves exhibit a clear jump that coincides with the abrupt increase of the G4 extension (figure \ref{fig:F_single_ch}a), so revealing the unfolding of the G4 structure. In those conditions, the unfolding patterns reveals a unique unfolding force $F_u$ that we define as the maximum force measured in the spring before the jump. Different to the atomistic simulations, we find that $F_u$ is in the same order of magnitude as the experimental values and that decreases when lowering the velocity. This behavior is due to the presence of thermal fluctuations, which facilitate the unfolding at lower pulling velocity. The nearly saturating behavior at low velocities indicates that the unfolding occur in the near equilibrium regime where $F_u$ is independent of the velocity. To express the velocity in real units, we need to specify the time units, which will be defined later when considering the mean value of the unfolding force as a function of the velocity. Figure~\ref{fig:F_single_ch}c) show the comparison between two simulations at $v=0.08$ and the experiment, showing  a clear agreement on the values of the unfolding force.

\begin{figure*}[!t]
\begin{center}
\includegraphics[width=0.25\textwidth,angle=-90]{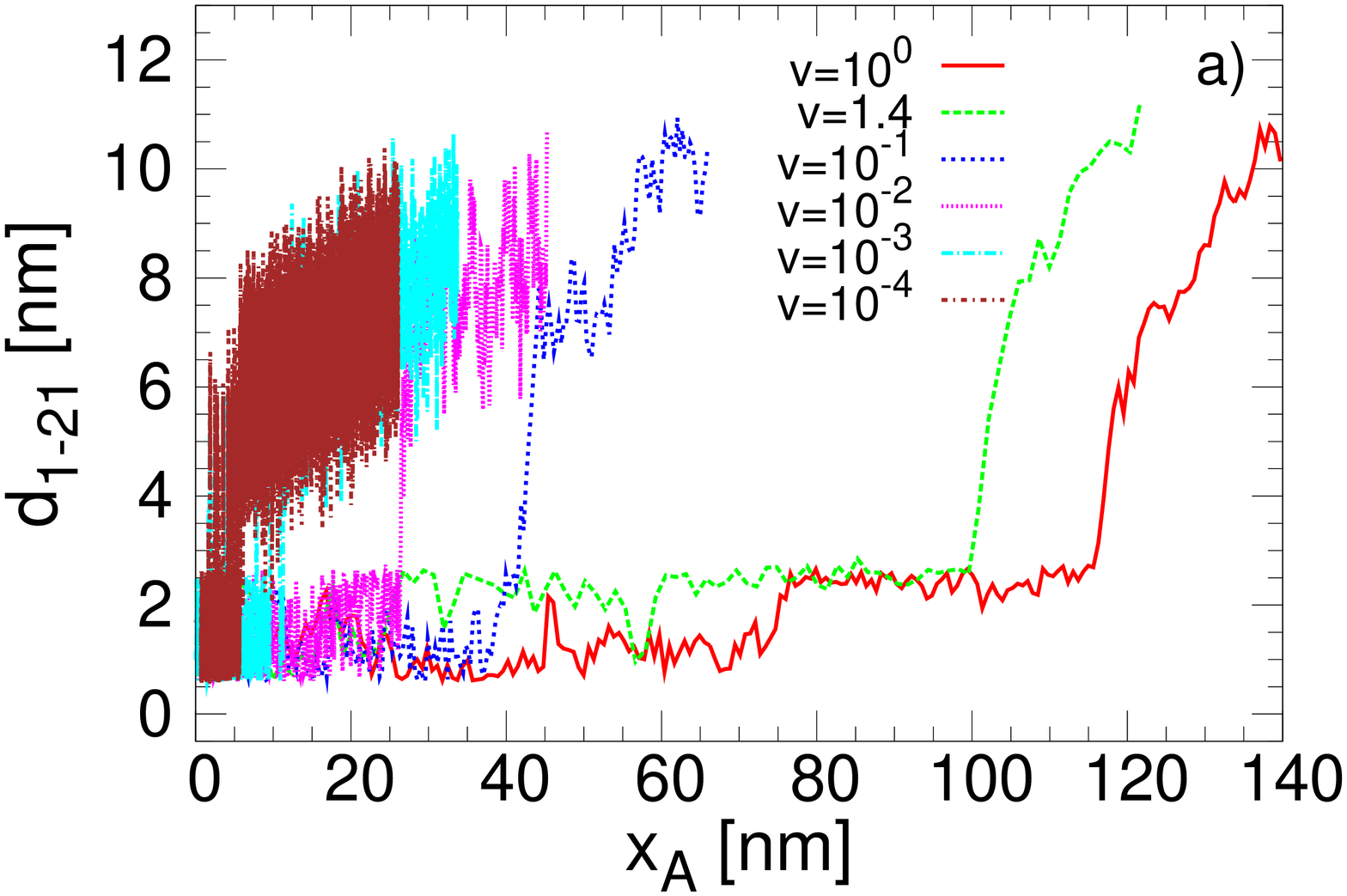}
\includegraphics[width=0.25\textwidth,angle=-90]{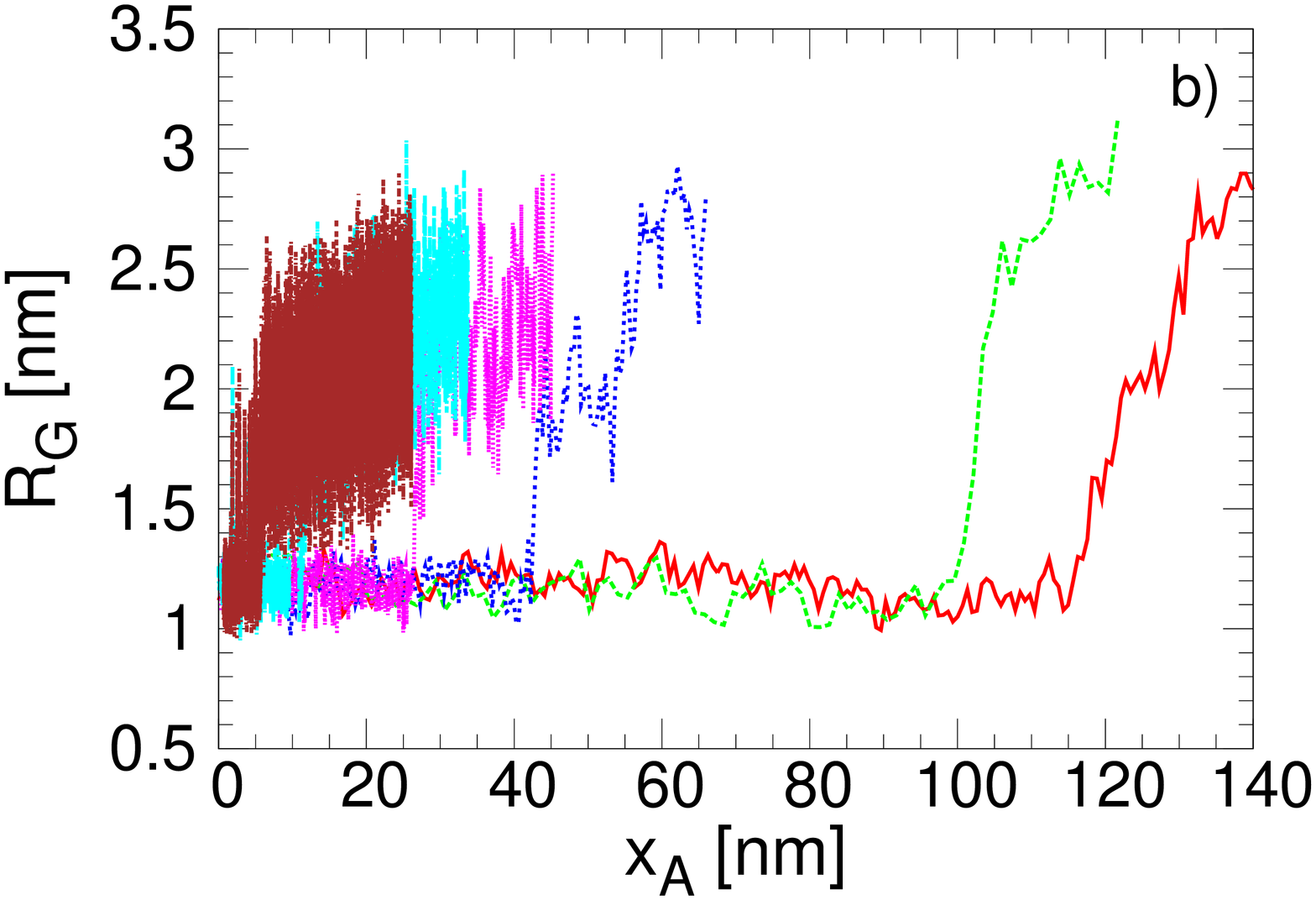}
\includegraphics[width=0.25\textwidth,angle=-90]{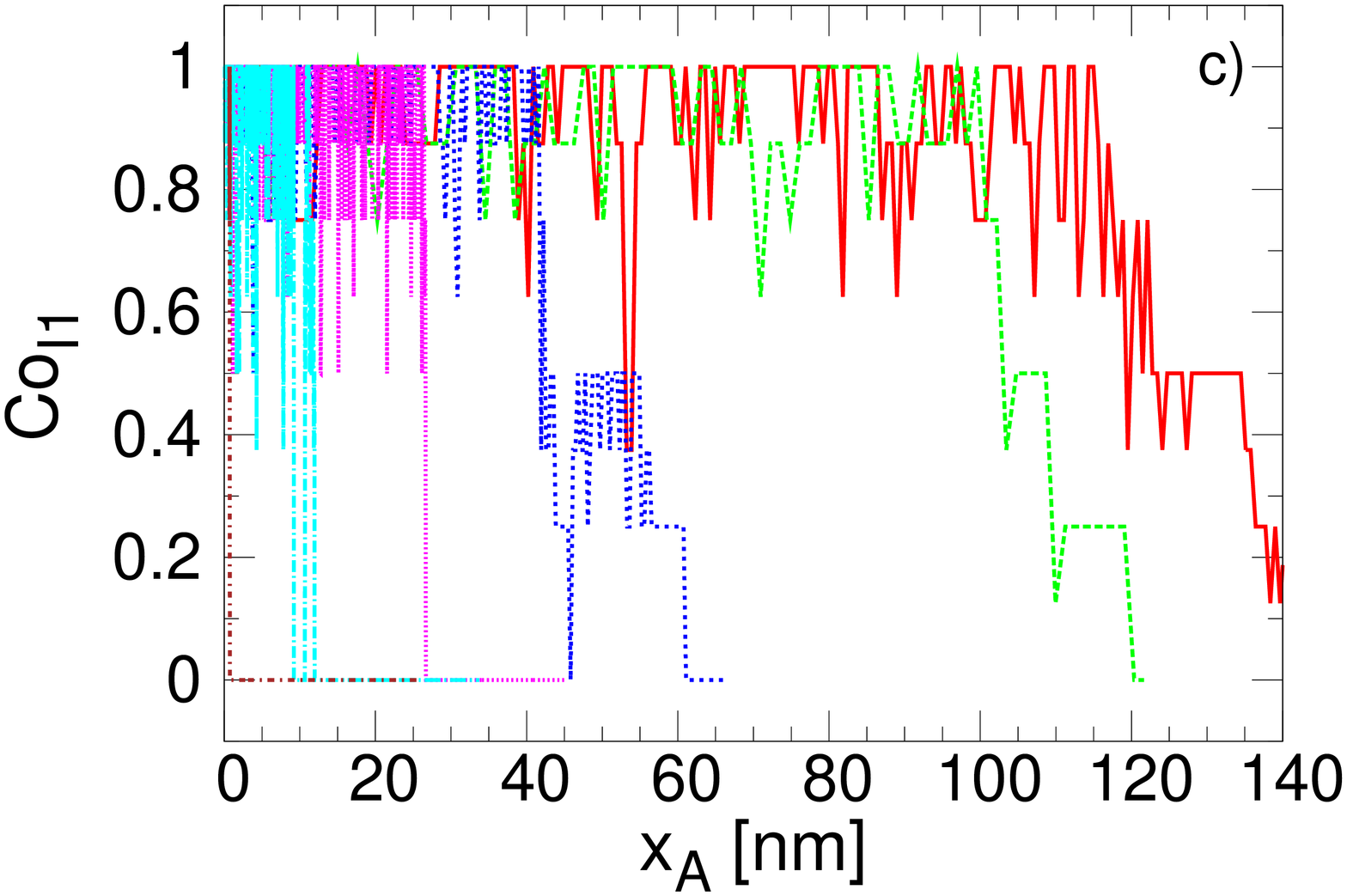}
\includegraphics[width=0.25\textwidth,angle=-90]{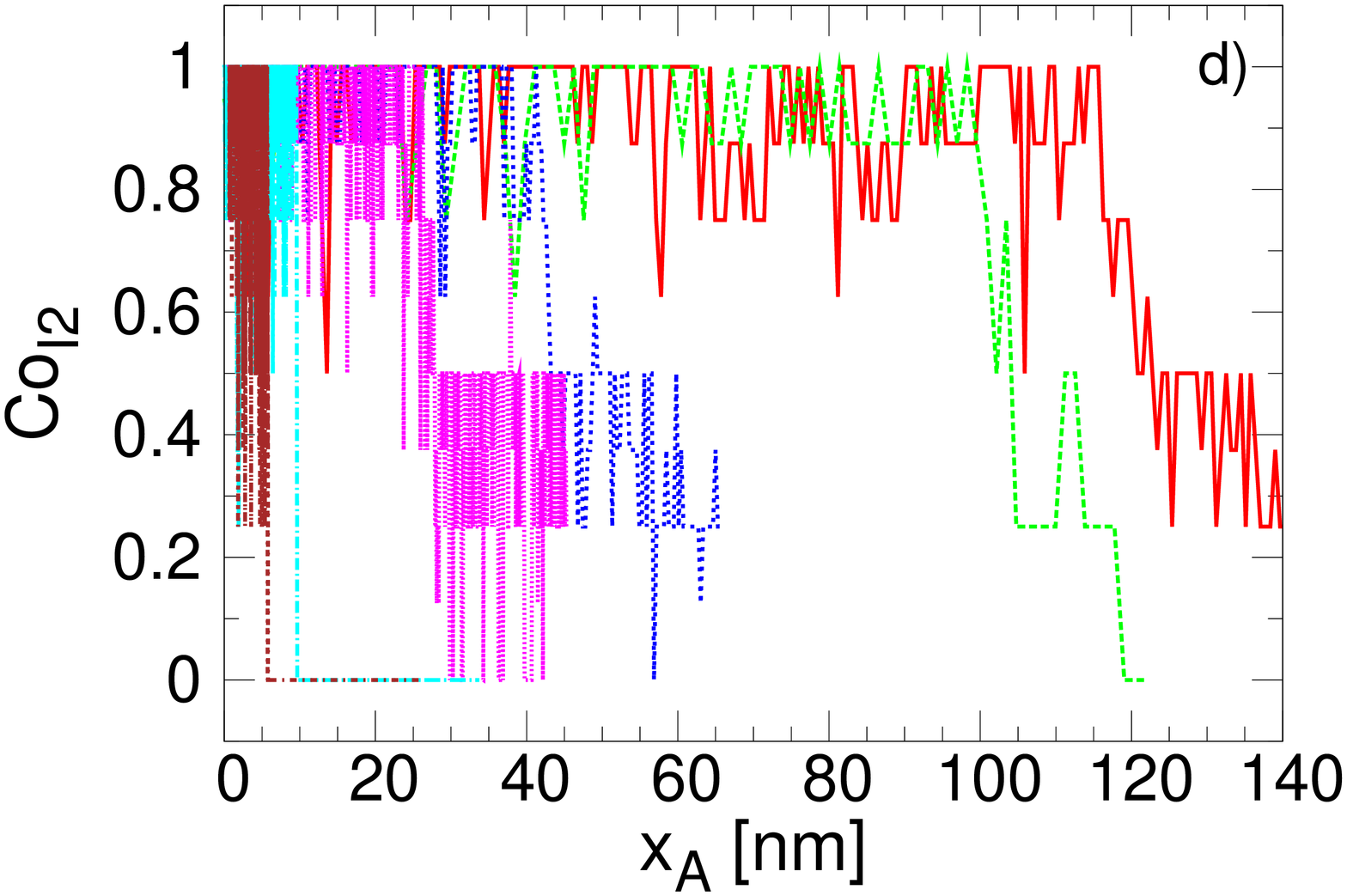}
\end{center}
\caption{Diagrams of the structural changes in the G4 as a function of the position of the pulling spring ($vt+x_0$) at different velocities. a) Distance between beads 1 and 21. b) Radius of gyration. c), d) Average coordination number of the first and second ion respectively with their neighbor guanines.
}
\label{fig:changes_single}
\end{figure*}

 Figure~\ref{fig:changes_single} shows different magnitudes that characterize the mechanical unfolding: the distance between beads 1 and 21  $d_{1-21}$ (panel a)), the gyration radius $R_G$ (panel b)) and the average coordination number between the two central ions and their eight neighbor guanines (panels c) and d)). We note that the mechanical unfolding pathway presents a similar behavior as the thermal one for all the simulated velocities. Specifically, the gyration radius $R_G$ increases abruptly almost at the same time as the ions coordination number goes to zero, showing the equivalence of the different measures. Moreover, the coperativity term in the Morse potential also presents a similar effect in the mechanical unfolding compared with the thermal unfolding: if the cooperativity is removed the unfolding occurs at higher values of the force and a multi-peak structure is observed in the force extension curves, corresponding to the consecutive rupture of the different planes.

\begin{figure}[!ht]
\begin{center}
\includegraphics[angle=-90,width=0.4\textwidth]{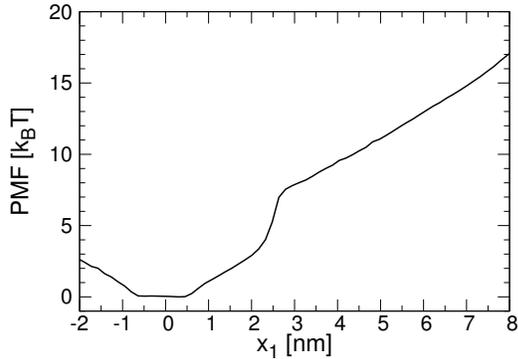}
\end{center}
\caption{Potential of Mean Force (PMF) calculated from umbrella sampling and the weighted histogram analysis method.}
\label{fig:W_single}
\end{figure}

Figure \ref{fig:W_single} shows the Potential of Mean Force (PMF). The PMF is the free energy along the extension $x_1$ and gives an equilibrium measure of the mechanical stability. It is calculated by using umbrella sampling \cite{Torrie1992} and the weighted histogram analysis method \cite{Kumar1992}. The initial conformations for calculating the PMF are taking from a pulling simulation at $v=0.01$. The PMF exhibit a change of the convexity around $x_1=2 \,\rm{nm}$ which is in correspondence with the extension at which the force jumps during the pulling simulations.

\begin{figure}[!t]
\begin{center}
\includegraphics[width=0.4\textwidth]{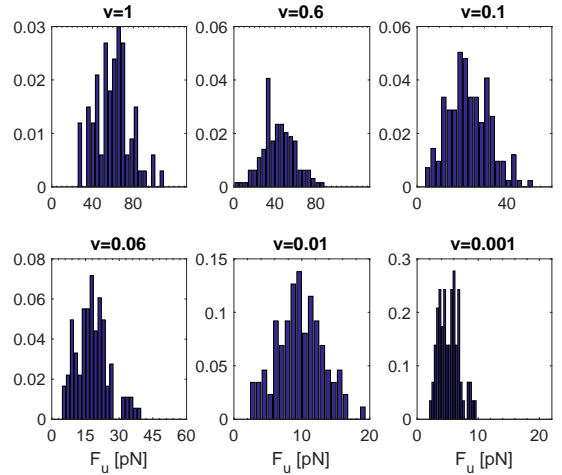}
\end{center}
\caption{Distribution of the unfolding force over 100 realizations at different velocities.}
\label{fig:F_hist}
\end{figure}

\begin{figure}[!t]
\begin{center}
\includegraphics[width=0.45\textwidth]{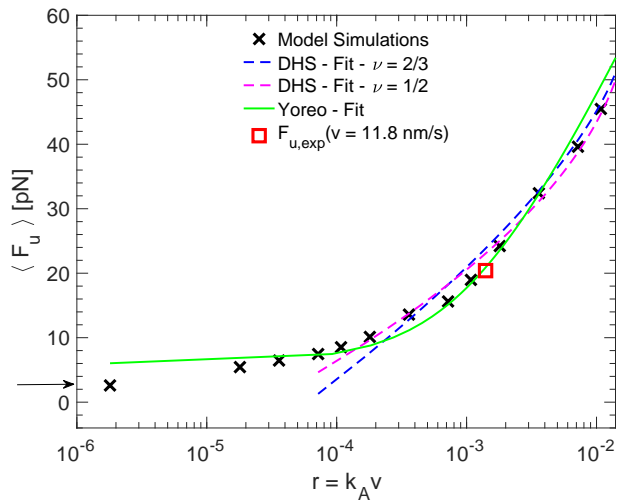}
\end{center}
\caption{Mean value of the unfolding force as a function of $r=k_Av$ and fitting to the kinetic models (note that the four lowest values of the loading rates are excluded from the fitting because in this region some rebinding events are observed during the unfolding, and then the kinetic models are not valid). Two experimental force values are included in this figure: $F_u=20.4 \,\rm{pN}$ from our constant velocity unfolding experiment (red square) and $F_u=2.5 \,\rm{pN}$ from the constant force experiment of Long et al. \cite{Long2013} (arrow in the left axis). The loading rate $r$ is expressed in dimensionless units. That is because we use its value at the experimental unfolding force to define the time units of our model (see text).}
\label{fig:F_v}
\end{figure}

To account for the influence of the stochastic effects during the unfolding, $N_R=100$ simulations at each velocity are performed. The distributions of the unfolding forces $F_u$ at each velocity are shown in figure \ref{fig:F_hist}. In agreement with the single realization results, the distributions displace towards lower values of the forces as the velocity is decreased. This behavior is better observed from the mean value of the unfolding force as a function of the velocity, or equivalently, as a function of the pulling rate $r=k_Av$ as shown in Fig.~\ref{fig:F_v}. In this figure we have included two experimental force values: the unfolding force $F_u=20.4 \,\rm{pN}$ obtained in the experiment at $v=11.8\,\rm{nm/s}$ and the equilibrium force $F_u=2.5 \,\rm{pN}$ obtained in the constant force experiment of Long et al. \cite{Long2013}. The loading rate in the dimensionless units of the mesoscopic model corresponding to this force value is $r$=0.0014, and the velocity is $v=r/k_A=0.0788 \,\rm{l_u/t_u}\approx 0.051 \,\rm{nm/t_u}$. Equaling this value to the experimental one $v=11.8 \,\rm{nm/s}$, we get the time unit to $t_u=0.0043\,\rm{s}$.

The dependence of $F_u$ \textit{vs} $r$ is similar to the observed in force dynamic spectroscopy experiments, where the strength of molecular bonds or the mechanical resistance upon unfolding is studied at different loading rates. Several analytical theories based on the Kramer's one-dimensional theory of diffusive barrier crossing in the presence of force have been proposed in order to get the kinetic parameters of a simplified energy landscape of the molecule at zero force from the $F_u$ \textit{vs} $r$ data: the height of the barrier separating the bounded/folded and unbounded/folded states $G^+$, the position of the barrier $x_u$ and the unfolding rate constant in the absence of force $k_0$. Two of the most widely used models in this analysis are the Bell-Evans-Ritchie (Bell) \cite{E. Evans1997} and the Dudko-Hummer-Szabo (DHS) \cite{Dudko2006}. The Bell model predicts a linear behavior for both the mean and the most probable rupture force as a function of the loading rate. In our data this linear behavior is not observed and we use, more consistently, the DHS model, which predicts a non linear behavior for high pulling velocity values. However, this model is not valid at low velocities where rebinding/refolding events can occur. For this reason we exclude the three lowest velocities from the fitting analysis with this model. Another kinetic model that takes into account the refolding events and then is valid in the region of low velocities is the Yoreo model \cite{Yoreo2012}. From this model the equilibrium unfolding force $f_{eq}$, which is independent of the pulling velocity, is obtained. $f_{eq}$ is the force at which the force dependent folding and unfolding rates are equal and depends on the elastic constant $k_A$ of the pulling spring: $f_{eq}=\sqrt{2k_AG^+}$. We will use this model to fit the simulated mean unfolding forces in the whole interval of loading rates.

The mean unfolding force as a function of the pulling rate with the DHS and Yoreo model reads, respectively:


\begin{equation}
\langle F_{u,D} \rangle=\frac{G^+}{\nu x_u} \left\lbrace 1- \left[\frac{1}{\beta G^+}  \ln \frac{k_0e^{\beta G^++\gamma}}{\beta x_u r}, \right]^{\nu} \right\rbrace
\label{eq:Dudko}
\end{equation}

\begin{equation}
\langle F_{u,Y} \rangle= f_{eq}+ \frac{k_BT}{x_u} e^{\left(\frac{k_u(f_{eq}k_BT)}{rx_u}\right)} E_1 \left(\dfrac{k_u(f_{eq}k_BT)}{rx_u} \right)
\label{eq:Yoreo}
\end{equation}

In the DHS model, eq. \ref{eq:Dudko}, $r$ is the rate of variation of the applied pulling force, $\gamma \approx 0.577$ is the Euler-Marchesoni constant and $\nu$ is a parameter that sets the shape of the free energy potential to a cusp potential (if $\nu=1/2$) or linear-cubic potential (if $\nu=2/3$). In the Yoreo model (see Eq.~\ref{eq:Yoreo}), $f_{eq}$ is the velocity independent equilibrium force and $E_1(z)=\int_z^{\infty} \frac{e^{-s}}{s} ds$ is the exponential integral which is approximated by $e^zE_1(z)\approx \ln (1+e^{-\gamma}/z)$. $k_u(f)=k_0\exp(\beta (fx_u-0.5k_Ax_u^2))$ is the force dependent unfolding rate from the Bell model. In our simulations the pulling is performed with a harmonic spring and then $r = dF(t)/dt = \frac{d}{dt}k_A(vt+x_1(0)-x_1(t)) \approx k_Av$.

The values of the parameters obtained from the fitting are summarized in table \ref{tab:Par}. Both variants of the DHS model, with $\nu=1/2$ and  $\nu=2/3$, give similar results in terms of the model parameters and the goodness of the fit. Differently, with the Yoreo model the estimated distance $x_u$ is lower than DHS, while the transition rate $k_0$ is higher. A similar trend has been observed in the unfolding of both Titin and RNA when these parameters are obtained using the Bell models, where lower values of $x_u$ and higher values of $k_0$ are obtained with respect to the DHS model \cite{Dudko2006}. This behavior agrees with the fact that both Yoreo and Bell models have the same dependence of the unbinding rate $k_u(F)$ as a function of the force. Force spectroscopy experiments performed in other G4 systems validate the order of magnitude of the parameters obtained with our model. For instance, Messieres et al. \cite{Mes2012} obtained $G^+=5.3 \, k_BT$, $x_u=0.9 \,\rm{nm}$ and $k_0=0.004 \,\rm{s^{-1}}$ for a parallel G4 with four guanine tetrads by simultaneously fitting the unfolding force distributions at different loading rates $r=2, 7, 24 \,\rm{pN/s}$ with the probability distribution function of the DHS model.

 \begin{table}
\caption{Fit parameters obtained from the mean unfolding force \textit{vs} the loading rate. $R^2$ is  the coefficient of determination that accounts for the goodness of the fitting. $k_A=0.4 \,\rm{pN/nm}$}
\begin{center}
\begin{tabular}{c c c c c}
\hline \hline
Model & $R^2$ & $G^+$ ($k_BT$) & $x_u$ (nm)& $k_0 (s^{-1})$  \\
\hline
Dudko $\nu=2/3$ & 0.97 & 5.59 &  0.61 &  0.017 \\
Dudko $\nu=1/2$ & 0.99 & 6.41 &  0.84 &   0.011\\
 Yoreo & 0.98 & 10.3 ($f_{eq}=6 \,\rm{pN}$) &  0.23 &  0.243\\
\hline
\hline
Exp. \\
\hline
Messiers $v$ const. \cite{Mes2012} & & 5.3 &  0.9 &  0.004 \\
Long $F$ const. \cite{Long2013}  & & ($f_{eq}=2.5 \,\rm{pN}$) &  0.6 &  0.24 \\
\hline
\hline
\end{tabular}
\end{center}
\label{tab:Par}
\end{table}

\subsection{Mechanical unfolding at T=0}

\begin{figure}[!ht]
\begin{center}
\includegraphics[width=0.3\textwidth,angle=-90]{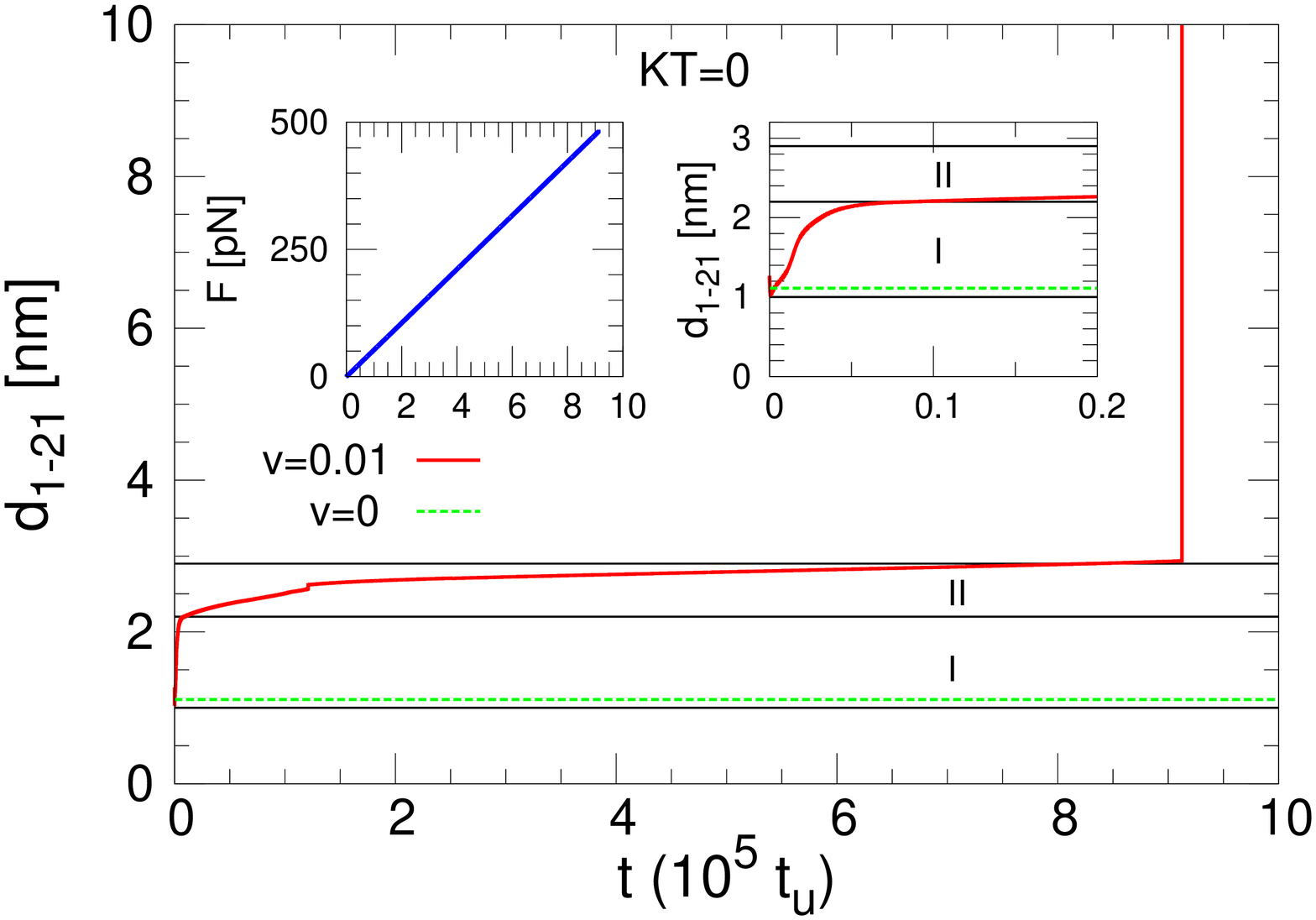}
\includegraphics[width=0.4\textwidth]{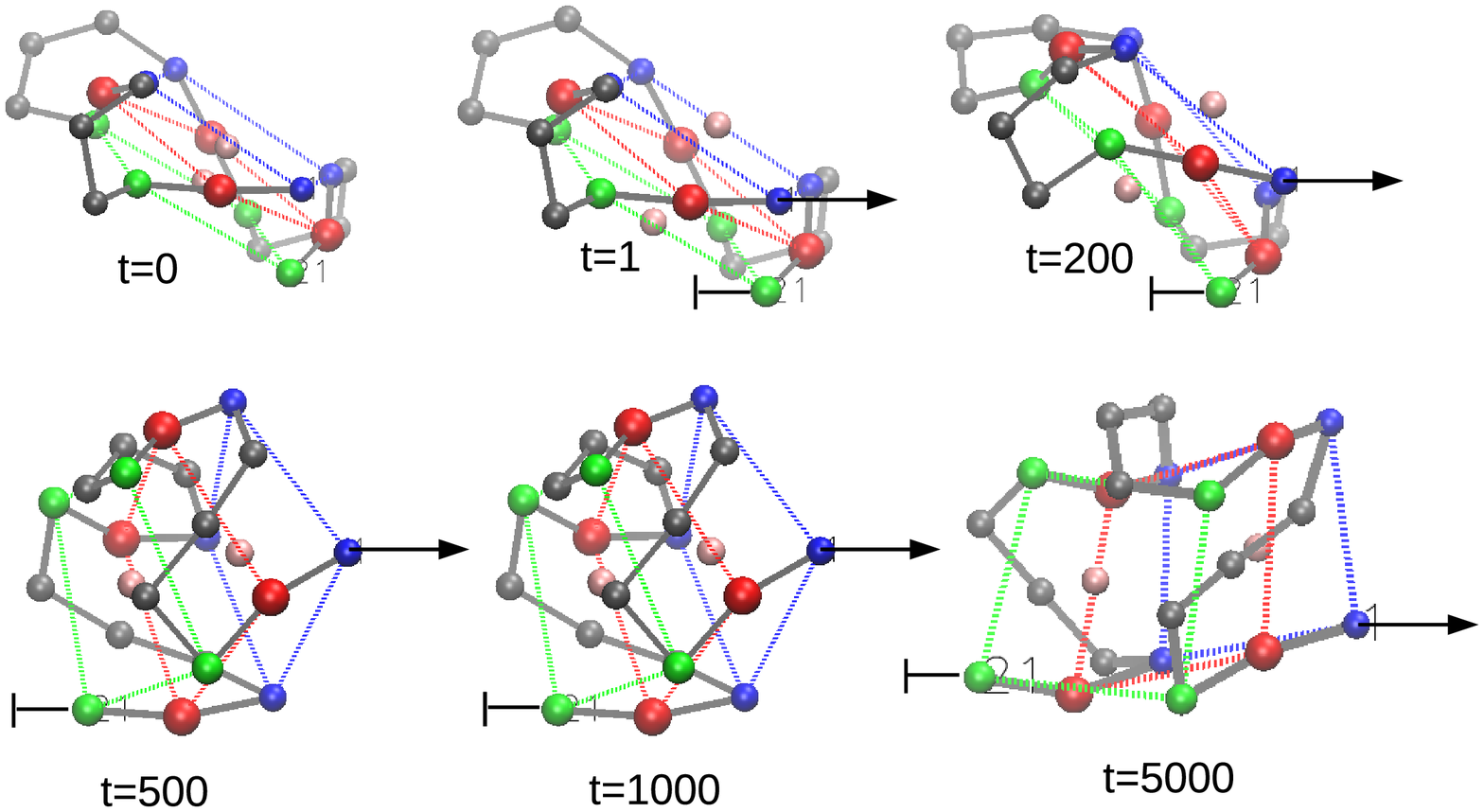}
\end{center}
\caption{Pulling simulation at $KT=0$. \textit{Top}: End to end distance $d_{1-21}$ and force $F$ as a function of the time. The dashed green lines mark the equilibrium value of $d_{1-21}$ at $v=0$. The solid black lines delimit the two distance intervals $1<d_{1-21}<2.2 \,\rm{nm}$ (I) and $2.2<d_{1-21}<2.9 \,\rm{nm}$ (II). \textit{Bottom}: Snapshots of the G4 during the pulling simulation. Beads belonging to the same plane have the same color (blue, red and green). Beads representing the bases of the loops and the two ions are colored in grey and pink, respectively.}
\label{fig:T0}
\end{figure}

To better understand the meaning of the parameters obtained from the fitting in the context of our model we look at one pulling simulation without thermal noise $KT=0$. Figure \ref{fig:T0} shows the behavior of  $d_{1-21}$ (distance between beads 1 and 21) and the force $F$ as a function of the time during a pulling simulation. We can identify different folded conformations before the unfolding occurrence at $t \approx 9 \times 10^5$. According to the behavior of $d_{1-21}$ as a function of time we can split the folded conformations in two main elongation ranges: i) $0.8<d_{1-21}<2.2 \,\rm{nm}$ ($t<0.05 \times 10^5$, visible in the right inset of the figure),  and ii) $2.2<d_{1-21}<2.9 \,\rm{nm}$ ($0.05 \times 10^5<t<9 \times 10^5$ in the main figure). When looking at the folded configurations corresponding to these two groups, we note that the increase in the extension for the first one is mainly related to a rearrangement of the distances and relative orientation between the planes, almost without difference in the distances of the very G4 planes: the plane has \emph{rotated} along their axes. In the second interval, the increase in the extension is mainly related with the \emph{stretching of the sides} of the planes. The possible extension of a guanine still bonded to the others in the G4 plane is ruled by the width of the Morse potential $\alpha^{-1}=0.065 \,\rm {nm} $. Looking at the $d_{1-21}$-values for different velocities at room temperature in figure \ref{fig:changes_single}a) we realize that the extension of the molecule before the unfolding depends on the velocity: we observe that for low velocities, the unfolding is more likely to occur when the G4 lies in configurations belonging to the first elongation subset $1<d_{1-21}<2.2 \,\rm{nm}$ and for higher velocities when belonging to  the second elongation subset $2.2<d_{1-21}<2.9 \,\rm{nm}$ ($0.05 \times 10^5<t<9 \times 10^5$. In fact, at low velocities ($v<0.01$) the system does not stabilize at lengths of the second $d_{1-21}$ subset, while, conversely, for $v>0.01$ the G4 does stabilize in the second subset before the G4 unfolds (see Fig.~\ref{fig:changes_single}a). This means that the unfolding pathway depends on the velocity, and particularly at low velocities, the dynamics is strongly assisted by the thermal fluctuations, with also the presence of refolding events. In these conditions the DHS theory does not work, and for this reason the points at the lowest velocities have been excluded from our fitting analysis. The parameters we got from the fitting are in agreement with the transition from the second folded subset of lengths to the unfolded state.

The values we calculated through the DHS model fit appear to be in agreement with the following conclusions: $x_u \approx 0.61 \,\rm{nm}$ for $\nu=2/3$ and  $x_u \approx 0.84 \,\rm{nm}$ for $\nu=1/2$ are both in the order of the expected extension length ($x_u=0.7 \,\rm{nm}$) of the second subset, which, in fact, is our estimation of the distance between the barrier and the state of the G4 before the unfolding. In addition, though the abrupt denaturation the G4 undergoes, the fit with the parameter $\nu=2/3$ reproduces a little better the expected barrier position, so suggesting that a smooth potential appears a better description than a cusp potential for this unfolding mechanism.
The values obtained for $G^+$ with the DHS model are lower than the free energy value obtained in the PMF at $x_1=2.2 \,\rm{nm}$ which appears to be the unfolding point, while the value obtained with the Yoreo model is closer to the value of the PMF in this point. These results corroborate the fact that the DHS model is describing the transition from the second subset length that requires less energy than the transition from the first substate, which is better described by the energy estimated from the Yoreo model. The equilibrium force $f_{eq}$ calculated from this model is however larger than the experimental value reported by Long et al. \cite{Long2013} which can be due to the fact that $f_{eq}$ depends on the elastic constant used for the pulling. Finally we note that the unfolding force obtained without the thermal contribution ($KT=0$) is $F_u(T=0) \approx 500 \,\rm{pN}$.

\section{Conclusions}
We have developed a mesoscopic model that captures the main features of DNA G4 thermal and mechanical stability. It is characterized by single beads representing each nucleotide of the ssDNA chain and the monovalent cations located at the central channel of the G4stem. Model parameters are obtained based on our previous atomistic study of telomeric DNA G4 \cite{Pupo2015}, on the melting temperatures of DNA G4 \cite{Phan1998}, and on the mechanical unfolding experiment conducted in this work.

Among the many potential terms in the model, the cooperativity term between the Guanines in the bond of the plane interactions has a focus role. In fact it allows the modification of the shape of the thermal transition from sharp -- when the cooperativity is strong -- to smooth -- when the cooperativity term is either low or removed. This phenomenological contribution can be adjusted to describe other conformations of the G4, such as the antiparallel or mixed arrangements of the strands.

The model correctly evidences the importance of the ions in the stabilization of the G4 structure, whose rupture force are increased with their presence.
More importantly, the model gives a very good description of the system under mechanical stretching. In this context, it is able to reproduce unfolding forces in the same order of magnitude as in experimental studies, which are impossible to reach with atomistic calculations. Related to that, we are able to explore wide time scales and study the unfolding pathways by using loading rates up to five orders of magnitude lower than those allowed in microscopic simulations. The evaluated mean rupture force as a function of the loading rate nicely reproduces the nonlinear increasing behavior observed in dynamic force spectroscopy experiments at high velocities and the almost force independent regime at low velocities, behaviors that are well fitted by the DHS and Yoreo models, respectively. The values of the parameters of the one dimensional free energy landscape assumed in the DHS and Yoreo models appears to be in very good agreement, on the one hand, with the umbrella sampling simulations (which permit to  determine the energy barrier $G^+$) and, on the other hand, to the pulling simulations at zero temperature (that allowed the estimation of the position of the potential barrier $x_u$). Moreover they are also in the order of magnitude of the corresponding parameters obtained in some other experiments \cite{Mes2012}.

The proposed model, eventually with some extension of it, can be used as a tool for performing systematic studies on mechanical stability of different G4 conformations. In this regard it may be applied to understand different G4 geometries according to the loop orientation (parallel vs antiparallel) or allow the mechano-chemical analysis of G4 tandem repeats~\cite{Garavis2013}, like those existing in human telomeric sequences.

\begin{acknowledgments}
Work supported by the Spanish Ministry of Economy and Competitiveness (MINECO), grant No.~FIS2014-55867, co-financed by FEDER funds. We also thank the support of the Arag\'on Government and Fondo Social Europeo to FENOL group.
Work in J.~R. A-G. laboratory was supported by a grant from MINECO,
No. MAT2015-71806-R).

\end{acknowledgments}

\end{document}